\documentclass[twocolumn,showpacs,preprintnumbers,floatfix]{revtex4}
\usepackage{graphicx}

\begin{document}


\title{The Kondo effect in C$_{60}$ single-molecule transistors}

\author{Lam H. Yu and Douglas Natelson}

\affiliation{Department of Physics and Astronomy, Rice University, 6100 Main St., Houston, TX 77005}

\date{\today}

\begin{abstract}

We have used an electromigration technique to fabricate C$_{60}$-based
single-molecule transistors.  We detail the process statistics
and the protocols used to infer the successful formation of a
single-molecule transistor.  At low temperatures each transistor acts
as a single-electron device in the Coulomb blockade regime.
Resonances in the differential conductance indicate vibrational
excitations consistent with a known mode of C$_{60}$.  In several
devices we observe conductance features characteristic of the Kondo
effect, a coherent many-body state comprising an unpaired spin on the
molecule coupled by exchange to the conduction electrons of the leads.
The inferred Kondo temperature typically exceeds 50~K, and signatures
of the vibrational modes persist into the Kondo regime.

\end{abstract}

\maketitle


A transistor with an active region consisting of a single small
molecule is the ultimate limit of the miniaturization of
three-terminal electronic devices.  Such single-molecule transistors
(SMTs) have been demonstrated using molecules of
C$_{60}$\cite{ParketAl00Nature}, C$_{140}$\cite{ParketAl03TSF}, cobalt
coordination complexes\cite{ParketAl02Nature}, and divanadium
complexes\cite{LiangetAl02Nature}.  These SMTs function as
single-electron transistors\cite{GrabertetAl92book,SohnetAl97}, with
conduction dominated by Coulomb blockade effects.  Because of the
small size of the molecules, charging energies and single-particle
level spacings in these devices are of the same order ($\sim$~1~eV), and
are much larger than those in semiconductor or metal single-electron
devices.  

Devices incorporating the latter two
molecules\cite{ParketAl02Nature,LiangetAl02Nature} exhibit signatures
of the Kondo effect\cite{Kondo64PTP} in their conduction properties.
In a single-electron device with a localized unpaired spin, it is
possible to observe Kondo physics, the growth of a correlated
many-body state comprising the localized spin interacting by
antiferromagetic exchange with the spins of the conduction electrons
of the leads.  This results in a Kondo resonance, an enhanced density
of states of the leads.  For dilute magnetic impurities in bulk metals
the result is enhanced scattering and therefore a low temperature rise
in the resistivity as the Kondo state develops.  Conversely in
single-electron
devices\cite{NgetAl88PRL,GlazmanetAl88JETPL,MeiretAl93PRL,WingreenetAl94PRB}
the result is enhanced transmission, manifested as a zero bias peak in
the conductance present when $T < T_{\rm K}$, the Kondo temperature.  The Kondo
energy scale $T_{\rm K}$ depends exponentially on the spin-lead coupling, and
on the energy of the localized level relative to the Fermi level of
the lead conduction electrons.  In single-electron devices, Kondo
physics has been observed in semiconductor quantum
dots\cite{GGetAl98Nature,CronenwettetAl98Science}, carbon
nanotubes\cite{NygardetAl00Nature,LiangetAl02PRL}, and the
single-molecule transistors incorporating metal ions mentioned above.
Kondo temperatures in these latter experiments range from 10~K to
25~K.

In this letter, we report measurements on SMTs incorporating
individual C$_{60}$ molecules coupled to gold source and drain electrodes.
We describe the fabrication procedure in detail, including the
statistics of the conduction properties of the resulting devices and
the protocols used to infer the successful formation of a SMT.  We
confirm the presence of C$_{60}$ vibrational resonances in devices in the
Coulomb blockade regime.  In several devices we report observations
consistent with Kondo physics.  The Kondo temperatures inferred from
the transport data typically exceed 50~K, significantly higher than
previously reported values in single-electron devices. The data also
suggest that signatures of inelastic vibrational processes can persist
well into the Kondo regime, evidence of coupling between a vibrational
excitation and the coherent many-body state.

\begin{figure}[h!]
\begin{center}
\includegraphics[clip, width=8cm]{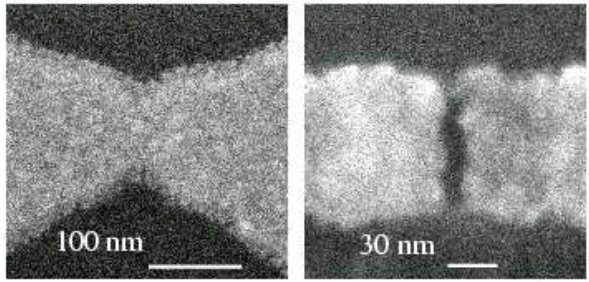}
\end{center}
\vspace{-5mm}
\caption{\small (a) Scanning electron image of a 1~nm Ti/15~nm Au 
constriction defined by electron beam lithography on 200~nm SiO$_{2}$
on a degenerately doped $p+$ Si substrate (gate electrode).  (b) A similar 
constriction after the electromigration procedure described in the
text.  The resulting interelectrode gap is less than 2~nm in this case.}
\label{fig:fig1}
\vspace{-5mm}
\end{figure}

The fabrication process is based on the electromigration
technique\cite{ParketAl99APL} employed in previous SMT
investigations\cite{ParketAl00Nature,ParketAl02Nature,LiangetAl02Nature,ParketAl03TSF}.
E-beam lithography (EBL) and lift-off processing are used to define
15-60 metal constrictions connected to contact pads on degenerately
doped $p+$ silicon substrates topped by 200 nm of thermal SiO$_{2}$.  An
example of such a constriction is shown in Fig.~\ref{fig:fig1}(a).
The metal, 1~nm of Ti (0.1~nm/s) and 15~nm of Au (0.2~nm/s), is
deposited in an e-beam evaporator with base pressure of $10^{-7}$~mB.
After liftoff the surface is cleaned by UV ozone for 5 minutes and O$_{2}$
plasma for 1 minute.  Then 80~$\mu$L of C$_{60}$ in toluene solution (1~mg
C$_{60}$ / 1~mL toluene) are spin cast (spin speed $\sim$ 900~RPM)
onto the array of junctions.  Scanning tunneling microscopy (STM)
images of C$_{60}$ deposited in this manner on an evaporated Ti/Au
film show approximately monolayer coverage of the metal by the
adsorbed molecules.  An indium contact to the $p+$ silicon substrate,
which serves as our back gate, is then made.  The C$_{60}$-decorated
junctions are place in a variable temperature vacuum probe station
(Desert Cryogenics) for the electromigration procedure and subsequent
electrical characterization.  The probe station is evacuated by
turbomolecular pump (base pressure of probe station at 300~K 
is $5 \times 10^{-5}$~mB) and cryopumped by a carbon felt 'sorption
pump thermally anchored to the incoming cryogen line.

A two-step variation of the electromigration technique is employed to
separate the constrictions into distinct source and drain electrodes.
Using an HP 4145B semiconductor parameter analyzer, at 300~K the
voltage across each junction is ramped from 0 to 400~mV while
monitoring the resulting current.  When electromigration-induced
junction breaking begins, indicated by decreased junction conductance,
the maximum voltage across the junction is reduced in steps of 40~mV
until the junction resistance is increased to 400-1000~$\Omega$.  Once
each junction is ``partially broken," the sample is cooled to liquid
helium temperatures.  At 4.2~K the electromigration process is
continued, with the maximum voltage ramps across the junction
increased in steps of 200~mV as the junction conductance decreases.
This process is halted when the resistance of the resulting electrodes
is $>$~100 k$\Omega$.  This two-step electromigration technique allows us
to make relatively high conductance electrodes consistently, which we
infer corresponds to a very small interelectrode gap.  Accurate scanning
electron microscopy (SEM) assessment of the interelectrode gap is
difficult because the newly exposed metal surfaces reconstruct when
the electrodes are warmed back to room temperature.  From SEM images
of these resulting electrodes (Fig.~\ref{fig:fig1}(b)), we can see
that the separation is less than 2~nm, the resolution of the SEM.
Since the diameter of a C$_{60}$ molecule is only 0.7~nm, closely
spaced electrodes are essential for C$_{60}$ SMTs.

Due to the stochastic nature of the electromigration process, every
electrode pair differs at the atomic scale.  Even if the closest
interelectrode separation is the right size, the presence of a
C$_{60}$ molecule at that location is probabilistic, and depends
strongly on the initial surface coverage of C$_{60}$.  If a molecule
is present, its couplings to the source, drain, and gate electrodes
are determined by the microscopic arrangement of the junction region,
which is different in every device.  We know of no atomic-scale
imaging technique at present that is capable of directly assessing
in-situ the presence of an individual molecule between source and
drain electrodes and the morphology of the gold electrodes adjacent to
the molecule.  Furthermore, the local charge environment of the SMT is
unknown {\it a priori} due to the existence of surface trap states
at the oxide surface.  However, by analyzing many samples
statistically we can estimate the percentage of the starting junctions
that will become SMT given our fabrication procedure.

From a sample size of 1094 electrode pairs created using the above
electromigration procedure, including control samples (no molecules;
different solvent exposures; different cleaning procedures), 70\% show
measurable source-drain currents after electromigration at 4.2~K.  We
have examined 475 junctions decorated with C$_{60}$ and
electromigrated as described above.  There are four classes of
conductance characteristics in the resulting electrode pairs.  (1) No
detectable source-drain current at $|V_{\rm SD}| = 0.1~V$ (34\%).  The
simplest explanation of these devices is that the breaking procedure
resulted in significantly too large a source-drain separation to
permit conduction. (2) Linear or slightly superlinear $I_{\rm
D}-V_{\rm SD}$ curves, consistent with simple tunneling behavior with
thermionic or field emission contributions at high bias (40\%).  The
most likely explanation for these devices is that no molecule is
present near the region of the source-drain gap that dominates
conduction. (3) Nontrivial $I_{\rm D}-V_{\rm SD}$ curves with steps
and abrupt changes in slope, but no detectable dependence on gate
voltage, $V_{\rm G}$ (15.3\%).  These devices likely have a molecule
or metal nanoparticle at the critical region of the interelectrode
gap, but local geometry screens the object from the effects of the
gate potential.  (4) Nontrivial $I_{\rm D}-V_{\rm SD}$ curves that may
be tuned significantly by varying $V_{\rm G}$ (10.7\%).  Conduction in
these devices may be examined as a function of $V_{\rm G}$,
$V_{\rm SD}$, and $T$, and compared with expectations for Coulomb
blockade dominated single-molecule transistors.  Occasionally some
electrode pairs that initially have linear or nongateable $I_{\rm
D}-V_{\rm SD}$ characteristics can change to exhibit interesting
conductance features upon thermal cycling to 300~K and back to 4.2~K.
Presumably this is results from a combination of molecular
rearrangement and metal reconstruction at 300~K.

In a single-molecule transistor in the Coulomb blockade regime, the
energetic cost of adding (removing) an electron to (from) the
molecule, given by the Coulomb charging energy of the molecule and the
energy difference between molecular levels, is sufficiently large, and
the coupling between molecules and leads is sufficiently poor, that at
most gate voltages the average charge on the molecule is fixed.  The
result of this charge quantization is a conductance gap, a region of
$V_{\rm SD}$ near zero bias where the conductance is suppressed.  The
energetic alignment of the molecular orbitals with respect to the
Fermi levels of the source and drain electrodes is determined by: the
work function of the metal; the electron affinity of the molecule; the
presence of any nearby charged defects or traps; and the capacitive
coupling to the gate electrode.  Because of this gate coupling, the
size of the conductance gap varies linearly and reversibly with
$V_{\rm G}$, since a more positive value of gate voltage makes it
energetically favorable to add an electron to the molecule.  At biases
larger than the conductance gap, conduction is permitted because the
source-drain potential difference is sufficient to overcome the
electron addition (subtraction) energy.  At certain values of gate
voltage (charge degeneracy points), it becomes energetically
degenerate for the charge of the molecule to change by one electron.
The result is that the conductance gap vanishes at zero bias, and as
$V_{\rm G}$ is increased through such a charge degeneracy point, the
average number of electrons on the molecule is increased by one.
Because of the extremely small size of the molecule, the Coulomb
charging energy of the molecule and the single-particle level spacing
are both large ($>>$~100~meV).

\begin{figure}[h!]
\begin{center}
\includegraphics[clip, width=8cm]{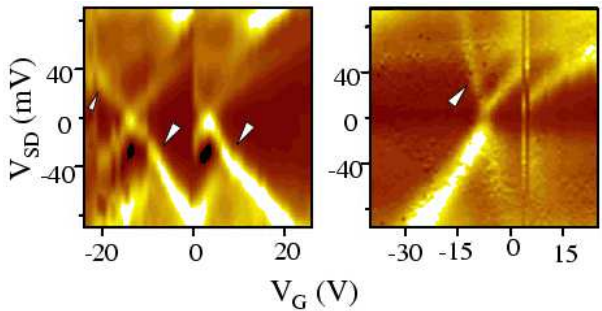}
\end{center}
\vspace{-5mm}
\caption{\small Color maps of the differential conductance $\partial I_{\rm D}/\partial V_{\rm SD}$ as a function of gate and source-drain voltage, for
two C$_{60}$ single-molecule transistors (black = 0 S; white (left) = $4 \times 10^{-6}$~S; white (right) = $1.2\times 10^{-5}$~S, $T = 4.2$~K).  The ``resetting'' of the
$V_{\rm G}$ axis is due to uncontrolled changes in the molecular charge
environment due to nearby traps in the oxide.  Moving from left to right 
along the $V_{\rm G}$ axis across the charge degeneracy point changes
the average charge on the molecule by one electron.  The arrows indicate
the 35~meV vibrational resonance previously reported in Ref.~\protect{\cite{ParketAl00Nature}}.}
\label{fig:fig2}
\vspace{-5mm}
\end{figure}

Previous investigations of single-molecule
transistors\cite{ParketAl00Nature,ParketAl02Nature,LiangetAl02Nature,ParketAl03TSF}
and other molecular devices\cite{ZhitenevetAl02PRL} have reported
additional resonances in the differential conductance at finite bias,
corresponding to excitations of molecular vibrational modes during the
transport process.  For C$_{60}$ devices\cite{ParketAl00Nature},
excitations of both the molecule-surface binding oscillation
($\sim$5~meV) and a mode intrinsic to the C$_{60}$ have been reported
($\sim$35~meV).  To characterize the electronic conduction in a SMT at
a fixed temperature, it is useful to plot the differential
conductance, $\partial I_{\rm D}/\partial V_{\rm SD}$, as a function
of $V_{\rm SD}$ and $V_{\rm G}$.  The data shown here were acquired by
measuring $I_{\rm D}$ as a function of $V_{\rm SD}$ for each value of
$V_{\rm G}$ using the HP~4145B semiconductor parameter analyzer.  The
differential conductance is obtained by numerically differentiation of
this data, using Savitzky-Golay smoothing to help reduce the noise in
the resulting curves at the cost of $V_{\rm SD}$ resolution.  For a
given temperature the resulting differential conductance is plotted as
a function of $V_{\rm SD}$ and $V_{\rm G}$ in a conductance contour
map, where the color of the map represents different levels of
conductance.

It is clear that not all gateable devices are C$_{60}$ SMTs because
about 5\% of control samples (with no molecules) show some
gateability.  Approximately half the gateable control samples have
clear Coulomb blockade style transport features, and the significant
majority of these are consistent (charging energy $\sim$~30~meV; many
accessible charge states) with metal islands left behind following
electromigration.  The existence of these metal islands has been
confirmed by SEM imaging.  However, in two control devices on one
substrate we observed Coulomb blockade features with charging energies
as large as 400~meV, though no apparent vibrational excited levels.
It is conceivable that some unintended adsorbed molecules contaminated
this particular set of devices during the preparation process.

We have developed criteria for deciding if a particular device with
nontrivial and gateable conductance is a SMT.  The existence of
Coulomb blockade is necessary but not sufficient.  The charging energy
of the Coulomb blockade feature must be relatively large, $>$~100~meV
(This is a challenging experimental requirement, because device
stability can be poor at such large biases.). The number of accessible
charge states should be reasonable, in light of solution-based
electrochemical redox information about the molecule.  Finally, the
existence of one or more vibrational resonances characteristic of the
molecule ({\it e.g.} 35~meV for C$_{60}$) in the conductance map is the
most indicative evidence that the device is a SMT.  The conductance
maps of two devices meeting these criteria are shown in Fig. 2.  Note
that the existence of surface trap states with charges that vary in
time leads to instability in some devices.  The random changes of the
local charge environment near the tips of the electrodes contribute to
some of the difficulties we have in observing the 35~meV vibrational
state.  In some of devices there appear to be multiple
Coulomb blockade features overlapping each other, possibly due to the
presence of multiple molecules between the electrodes.

\begin{figure}[h!]
\begin{center}
\includegraphics[clip, width=8cm]{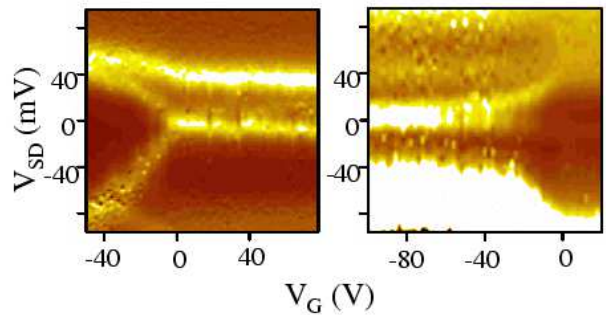}
\end{center}
\vspace{-5mm}
\caption{\small Conductance maps for two devices showing apparent 
Kondo behavior, manifested as the appearance of a zero bias conductance
peak in one molecular charge state.  Note the presence of $\sim$ 35~meV
sidebands persisting into the Kondo regime.  (dark = 0 S; white (left) = $1.5 \times 10^{-5}$~S; white (right) = $6 \times 10^{-7}$~S; $T=4.2$~K) }
\label{fig:fig3}
\vspace{-5mm}
\end{figure}

Four of the devices successfully electromigrated at 4.2 K exhibit data
like that shown in Fig. 3.  When a transition is made from one charge
state to another, a zero bias resonance appears in the differential
conductance.  Although a zero bias resonance is occasionally seen in
control samples, the feature is always gate independent.  The
transitional behavior from Coulomb blockade to zero bias resonance is
{\it never} observed in devices made without C$_{60}$ molecules.  This
transition is consistent with the Kondo effect in single-electron
devices, where the Kondo resonance can only exist when the active
region of the device has an odd number of electrons.

When describing Kondo phenomena in single-electron devices, the width
of the localized state, $\Gamma$, is defined as the sum of the level
widths due to the couplings, $\Gamma_{\rm S}, \Gamma_{\rm D}$ of the
localized state to the source and drain, respectively.  The energy
difference between the localized state (tunable by gate voltage) and
the Fermi level of the leads is $-\epsilon$.  When $0 < \epsilon/\Gamma <
1$, the system is said to be in the ``mixed-valence'' regime, while
$\epsilon/\Gamma >> 1$ corresponds to the Kondo regime.  In this
limit, the Kondo temperature is given\cite{Haldane78PRL,GGetAl98PRL}
by $T_{\rm K} = 0.5 (\Gamma U)^{1/2}\exp(-\pi \epsilon/\Gamma)$.

Note that $T_{\rm K}$ depends {\it exponentially} on $\Gamma$, and in
a single-molecule transistor $\Gamma$ depends {\it exponentially} on
the relative position of the molecule with respect to the leads.  This
steep dependence has made it extremely challenging to examine this
physics over a large temperature range, due to temporal instability of
the molecule-metal configuration.  For example, in one device of the
type shown in Fig.~\ref{fig:fig3}, while acquiring conductance data
the device switched irreversibly from exhibiting a Kondo-like zero
bias peak, as shown, to standard Coulomb blockade of the type shown in
Fig.~\ref{fig:fig2}, without a change in the charge degeneracy point.
Within the Kondo picture, this change corresponds to the molecule-lead
coupling changing to lower $T_{\rm K}$ below $T$.  We have been able
to acquire data over a limited temperature range for two devices
exhibiting the Kondo-like resonance.

To analyze the zero bias resonance data in the context of Kondo
physics, we follow previous SMT Kondo
investigations\cite{ParketAl02Nature,LiangetAl02Nature}.  The Kondo
temperature may be inferred in two different ways.  First, assuming
spin-1/2, at fixed gate voltage the zero bias conductance $G$ may be
monitored as a function of temperature, and fit with the formula
$G(T)=G_{0}/(1+2^{1/s}T^{2}/T_{\rm K}^{2})^{s}$, where $G_{0}$ is a
constant and $s \sim 0.22$ in the Kondo regime\cite{GGetAl98PRL}.  In
both devices mentioned above, $G(T)$ is nearly constant, decreasing only
slightly from 4.2~K up to $\sim$~30~K. This is consistent with large
Kondo temperatures, $T_{\rm K} >$~100~K, though the data do not put an
upper bound on $T_{\rm K}$.

\begin{figure}[h!]
\begin{center}
\includegraphics[clip, width=8cm]{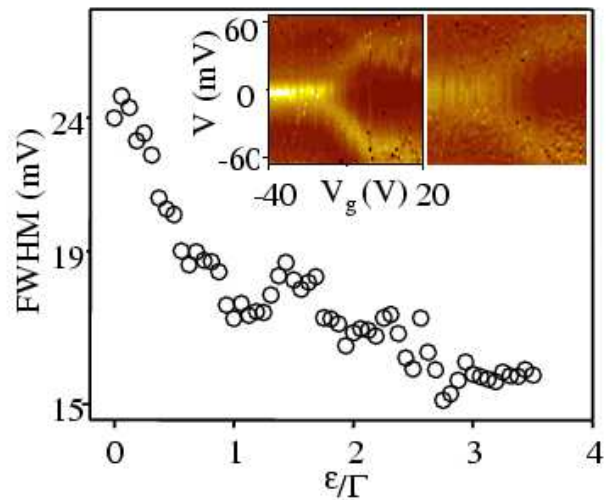}
\end{center}
\vspace{-5mm}
\caption{\small A plot of full-width at half maximum of the zero bias
conductance peak as a function of $\epsilon/\Gamma \propto |V_{\rm G}-V_{\rm c}|$, where $V_{\rm c}$ is the gate voltage of the charge degeneracy
point.  Inset:  the conductance map of this device at 18~K (left) and 50~K (right),
showing the persistance of the zero bias peak to quite high temperatures (dark = 0~S; white = $1.15 \times 10^{-5}$~S).}
\label{fig:fig4}
\vspace{-5mm}
\end{figure}

Further analysis is possible making use of the conductance maps.
Consider the device shown in Fig.~\ref{fig:fig4}.  The slopes of the
Coulomb blockade gap edges approximately give $\alpha$, the conversion
factor between gate voltage and the source-drain bias energy scale.
For this sample, $\alpha \approx 2$~meV/$V_{\rm G}$, not surprising
given the 200~nm thickness of the gate oxide.  As $T \rightarrow 0$,
the width of the zero bias Coulomb blockade conductance peak as a
function of gate voltage saturates to $\Gamma\approx32$~meV.  Using
$\alpha$, and knowing that $\epsilon = 0$ at the charge degeneracy
point, we can find $\epsilon(V_{\rm G})$.  In a Kondo device the
full-width at half maximum of the zero bias conductance peak is
expected to
be\cite{GGetAl98Nature,CronenwettetAl98Science,vanderwieletAl00Science,NygardetAl00Nature,ParketAl02Nature,LiangetAl02Nature}
$\sim k_{\rm B}T_{\rm K}/e$.  Fig.~\ref{fig:fig4} shows this FWHM as a
function of $\epsilon/\Gamma$ for that particular device.  As with
$G(T)$, this data is, within the noise, nearly temperature independent
below 20~K, and an average of several low temperatures is plotted.
This FWHM increase as $\epsilon \rightarrow 0$ is consistent with the
Kondo behavior reported by Liang {\it et al.}\cite{LiangetAl02Nature}.
The high effective Kondo temperatures implied by this data are
clear from the insets to Fig.~\ref{fig:fig4}, which demonstrate
that the zero bias resonance is still visible up to at least 50~K.

Other studies of the Kondo effect in single-molecule
devices\cite{ParketAl02Nature,LiangetAl02Nature} were able to
demonstrate the Zeeman splitting of the Kondo resonance in an
applied magnetic field. 
Because of the large intrinsic width ($\sim$ 10-20~meV) of the
zero bias resonance (consistent with high Kondo temperatures)
that we observe, and the small size of the Zeeman splitting
($115~\mu$eV per Tesla for a free electron), this would
be extremely difficult to observe in our samples. 

We also note that sidebands appear in these samples parallel to the
zero bias peak.  In the two samples shown in Fig.~\ref{fig:fig3}, the
sidebands are located at $V_{\rm SD}\sim$ 35~meV, and appear to evolve
from the inelastic resonances ascribed to vibrational excitations
during the tunneling process.  Hints of this behavior were noted in
Ref.~\cite{LiangetAl02Nature}, but here there is coincidence of the
sideband voltage and the known molecular vibrational resonance.
Conductance under these nonequilibrium conditions would involve some
interplay between inelastic processes and the coherent many-body Kondo
state.

We have successfully created C$_{60}$-based single-molecule
transistors using the electromigration technique\cite{ParketAl99APL}.
Statistics on success are comparable to those reported by other
investigators, and have been presented in detail, along with a
discussion of controls and protocols.  We find evidence of Kondo
physics with large Kondo temperatures in some of these SMTs without
having a metal ion present in the molecule.  As mentioned above,
sideband resonances indicate conduction processes that involve both
many-body correlations and inelastic coupling to vibrational modes.
Finally, we note that a high temperature Kondo resonance in C$_{60}$
adsorbed on a noble metal electrodes would explain the surprisingly
narrow local density of states observed in scanning tunneling
microscopy experiments with C$_{60}$ tips\cite{KellyetAl96Science}.

The authors gratefully acknowledge the support of the Robert A. Welch
Foundation, the Research Corporation, and the David and Lucille
Packard Foundation.  The authors also thank K. Kelly and A. Osgood for
STM characterization, and P.L. McEuen, P. Nordlander, H. Park,
D.C. Ralph, A. Rimberg, J.M. Tour, and R.L. Willett for useful
conversations.


\end{document}